\input harvmac

\def\p{\partial}

\def\half{{1\over 2}}

\Title{hep-th/0001193}{\vbox{\centerline{A Note On Relation Between Holographic
RG Equation
}
\vskip15pt
\centerline{And Polchinski's RG Equation
}}}
\vskip20pt

\centerline{Miao Li}
\vskip 10pt
\centerline{\it Institute of Theoretical Physics}
\centerline{\it Academia Sinica}
\centerline{\it Beijing 100080} 
\centerline { and}
\centerline{\it Department of Physics}
\centerline{\it National Taiwan University}
\centerline{\it Taipei 106, Taiwan}
\centerline{\tt mli@phys.ntu.edu.tw}

\bigskip

We clarify the relation between the recently formulated holographic
renormalization group equation and Polchinski's exact renormalization
group equation.

\Date{Jan. 2000}

\nref\dvv{J. de Boer, E. Verlinde and H. Verlinde,
{\it On the Holographic Renormalization Group}, hep-th/9912012;
E. Verlinde and H. Verlinde, {\it Gravity and the Cosmological
Constant}, hep-th/9912018.}
\nref\kv{J. Khoury and H. Verlinde, {\it On Open/Closed String
Duality}, hep-th/0001056.}
\nref\ak{E.T. Akhmedov, {\it A Remark on the
 AdS/CFT Correspondence and the 
 Renormalization Group  Flow},
 Phys. Lett. {\bf B442} (1998) 152,
 hep-th/9806217.}
\nref\alg{E. Alvarez and C. G\'omez
 {\it Geometric Holography, the Renormalization
 Group and the c-Theorem}, Nucl. Phys. 
{\bf B541} (1999) 441, hep-th/9807226.}
\nref\fgpw{ D. Z. Freedman, S. S. Gubser, K. Pilch, N. P. Warner,
{\it Renormalization Group Flows from Holography--Supersymmetry and a 
c-Theorem}, hep-th/9906194.}
\nref\ita{L. Girardello, M. Petrini, M. Porrati, A. Zaffaroni, 
{\it The Supergravity Dual of $N=1$ Super Yang-Mills Theory}, hep-th/9909047;
{\it Novel Local CFT and Exact Results on Perturbations of N=4 Super Yang 
Mills from AdS Dynamics}, hep-th/9810126;
M. Porrati, A. Starinets, {\it RG Fixed Points in Supergravity Duals of 
4-d Field Theory and Asymptotically AdS Spaces}, Phys.Lett. 
{\bf B454} (1999) 77, hep-th/9903241.}
\nref\per{V. Balasubramanian, P. Kraus, {\it Space-time and the Holographic 
Renormalization Group}, Phys.Rev.Lett 83 (1999) 3605, 
hep-th/9903190.}
\nref\st{K. Skenderis, P.K. Townsend, 
{\it Gravitational Stability and 
Renormalization-Group Flow}, hep-th/9909070.}
\nref\mit{ O. DeWolfe, D.Z. Freedman, S.S. Gubser, A. Karch, {\it Modelling t
he fifth dimension with scalars and gravity}, hep-th/9909134.}
\nref\smid{C. Schmidhuber, {\it AdS-Flows and Weyl Gravity}, 
hep-th/9912155.}
\nref\joe{J. Polchinski, {\it Renormalization and Effective Lagrangians},
Nucl. Phys. B231 (1984) 269.}
\nref\ads{J. Maldacena, {\it The Large N Limit of Superconformal
Field Theories and Supergravity}, Adv. Theor. Math. Phys. 2 (1998) 231,
hep-th/9711200; E. Witten, {\it Anti-de Sitter Space and Holography},
Adv. Theor. Math. Phys. 2 (1998) 253,
hep-th/9802150; S. Gubser, I. Klebanov and A. Polyakov, {\it Gauge Theory
Correlators from Noncritical String Theory}, Phys. Lett.
B428 (1998) 105, hep-th/9802109.}
\nref\mm{E. Brezin and S. Wadia, {\it The large N expansion in quantum 
field theory and statistical physics: from spin systems to two-dimensional 
gravity}, Singapore, World Scientific (1993).}
\nref\ty{T. Yoneya, {\it Action Principle, Virasoro Structure and
Analyticity in Nonperturbative Two-dimensional Gravity}, Int. J. Mod.
Phys. A7 (1992) 4015.}
\nref\morris{T. R. Morris, {\it A Manifestly Gauge Invariant Exact 
Renormalization Group}, hep-th/9810104.} 
\nref\hirano{S. Hirano, {\it Exact Renormalization Group and Loop Equation},
hep-th/9910256.}
\nref\pw{G. Parisi and Y.S. Wu, Sci. Sin. 24 (1981) 484.}
\nref\vipul{V. Periwal, {\it String field theory Hamiltonian from Yang-Mills
theories}, hep-th/9906052; G. Lifschytz and V. Periwal, {\it Dynamical
Truncation of the String Spectrum at Finite N}, hep-th/9909152.}
\nref\lw{M. Li and Y.S. Wu, {\it Holography and Noncommutative Yang-Mills},
hep-th/9909085.}

The holographic renormalization group flow has been clarified in the context of
AdS/CFT correspondence \dvv\ as well as in the context of
open string versus closed string \kv. (For earlier attempts in this,
see \refs{\ak--\mit}, also see \smid.) It was noticed in these papers
that the holographic RG equation arising either from the Hamilton-Jacobi
theory of supergravity or from world-sheet considerations has a strong
resemblance to Polchinski's exact RG equation \joe. 

Apparently the
two equations are different. We aim in this short note to clarify the
relation between them.
The effective action in the AdS/CFT correspondence \ads\ is defined
as a functional of coupling constants, while the effective action
of Polchinski is a functional of the fundamental fields. Thus the holographic
RG equation is naturally a differential equation in coupling coupling
constants, and the Polchinski RG equation is one in fundamental
fields. 

For simplicity and without loss of generality, we will consider
the field theory of a single Hermitian matrix. 
We start with a single
matrix model in 0 dimension to illustrate some of our ideas. 
Although in the 0-dimensional matrix model
there is no infinity to remove, one still can design an artificial
RG flow by introducing a ``cut-off'' in the quadratic term in the
action
\eqn\bare{S_0 =-\half NK(t)\tr \Phi^2,}
where $\Phi$ is a Hermitian matrix, $K(t)$ is the cut-off propagator,
depending on $t=\ln a$, $a$ is the running cut-off. Unlike in a genuine
field theory, where
once an interaction term is introduced in the action, many other
terms will be generated with a nontrivial $K(t)$ in order to keep the
physics invariant under changing $t$. In our case, there is much
freedom in satisfying the RG flow, as we shall explain later. 
To mimic the ${\cal N}=4$
super Yang Mills theory, we introduce the interaction part as a sum
of single trace operators
\eqn\inte{S_1=N\sum_{n\ge 3} \phi_n(t)\tr\Phi^n,}
where again a factor N is introduced to follow the usual large N
field theory convention. With this convention, the effective action
as a functional of $\phi_n$ defined by
\eqn\effc{e^{S(\phi(t),t)}=\int[d\Phi]e^{S_0+S_1},}
has the usual genus expansion
\eqn\gexp{S(\phi_n)=\sum N^{2-2h}F_h,}
and the connected two point functions of operators $\tr\Phi^n$
in the leading order is proportional to $N^0$. 
We pause to emphasize that it is crucial
to introduce only single trace operators. In the AdS/CFT correspondence,
a single trace operator is related to a field in SUGRA or string theory,
the role of which is played by $\phi_n$ in our toy model. Multi-trace
operators are related to multi-particle states. The RG equation as 
formulated in \dvv\ has to do with only single trace operators.

To ensure the effective action $S$ be independent of $t$, $S_1$ must
satisfy a differential equation \joe. As we shall see, this differential
equation can not be satisfied by our model in which $S_1$ contains
only single trace operators. Thus we need to relax this equation to be
the one valid only when taken average in the path integral, namely
\eqn\ave{\langle \p_t S_1+K_1^{ij,lk}{\p S_1\over\p \Phi_{ij}}
{\p S_1\over\p \Phi_{lk}}+K_2^{ij,lk}{\p^2 S_1\over \p\Phi_{ij}\p\Phi_{lk}}
+K_3\rangle=0,}
where 
$$\langle {\cal O}\rangle={\int[d\Phi]{\cal O}e^{S_0+S_1}\over
\int[d\Phi]e^{S_0+S_1}}.$$
we will determine $K_1, K_2, K_3$ momentarily. Note that the constant
term $K_3$ can be removed by a shift of $S_1$, and this shift can be
absorbed into the definition for the measure of $\Phi$. Thus in
the following we will ignore this term.

For completeness, we will derive eq.\ave. We start with
\eqn\sta{\p_t e^S=\int [d\Phi](-\half\p_tK\tr\Phi^2+\p_tS_1)e^{S_0+S_1}=0.}
Use \ave\ to replace $\p_t S_1$ in \sta. Next, use the fact
$$\eqalign{&\int [d\Phi]K^{ij,lk}_1{\p S_1\over \p\Phi_{ij}}{\p S_1
\over \p\Phi_{lk}}
e^{S_0+S_1}=\int [d\Phi]K^{ij,lk}_1{\p S_1\over\p\Phi_{ij}}
(NK\Phi_{kl}+{\p\over \p\Phi_{lk}})e^{S_0+S_1}\cr
&=\int [d\Phi]\left( NKK^{ij,lk}_1{\p S_1\over\p\Phi_{ij}}\Phi_{kl}
-K^{ij,lk}_1{\p^2 S_1\over \p\Phi_{ij}\p\Phi_{lk}}\right)
e^{S_0+S_1},}$$
we see that if we choose $K_2^{ij,lk}=K^{ij,lk}_1$, then the second
derivatives of $S_1$ cancel. Applying the same trick to the
first term in the last line of the
above equation 
$$\eqalign{&\int [d\Phi]NKK^{ij,lk}_1\Phi_{kl}\left(NK\Phi_{ji}+{\p\over
\p\Phi_{ij}}\right)e^{S_0+S_1}\cr
&=\int [d\Phi](N^2K^2K^{ij,lk}_1\Phi_{ji}\Phi_{kl}-NK_1^{ij,ji})
e^{S_0+S_1}.}$$
Now the first term in the above can be used to cancel the first term
in \sta\ if 
\eqn\ans{K_1^{ij,lk}=\half N^{-1}\p_tK^{-1}\delta_{ik}\delta_{jl}.}
And the inhomogeneous term is removed by choosing $K_3=-\half
N^2\p_t\ln K$. However, as we remarked before, this term can be
absorbed into a redefinition of the measure and henceforth we will ignore
it. To summarize, we have derived the following equation
\eqn\wjoe{\langle \p_t S_1+\half N^{-1}\p_tK^{-1}
\left({\p S_1\over \p\Phi_{ij}}{\p S_1\over \p\Phi_{ji}}
+{\p^2 S_1\over\p\Phi_{ij}\p\Phi_{ji}}\right)\rangle=0.}
As we advertised, this is the weak form of Polchinski's equation.

The above equation is not valid if the average symbol is removed.
(We call this equation the strong form of Polchinski equation) 
To see this, we compute
\eqn\oned{{\p S_1\over \p\Phi_{ij}}{\p S_1\over \p \Phi_{ji}}=
=N^2\sum_{n\ge 4} g_n\tr\Phi^n,}
where
\eqn\gn{g_n=\sum_l l(n+2-l)\phi_l\phi_{n+2-l}.}
And 
\eqn\twod{{\p^2 S_1\over \p\Phi_{ij}\p\Phi_{ji}}=\sum
NG_{mn}\tr\Phi^m\tr\Phi^n}
with
\eqn\gmn{G_{mn}=(m+n+2)\phi_{m+n+2}.}

If the original Polchinski equation applies, then
$\p_t S_1$ contains only single trace operators, and can be used
to balance the single trace operators in \oned. However, \twod\
contains double trace operators, and can not be balanced in Polchinski
equation, in the large N limit, since these operators are
new independent operators. In order to solve Polchinski equation, 
we need to introduce in $S_1$ double trace operators. This in turn generates
triple trace operators in $\p^2 S_1$, etc. Thus in order for the Polchinski
equation to hold, all multiple trace operators must be introduced.
With a little thought, it is easy to realize that this conclusion
holds for any matrix model, including ${\cal N}=4$ SYM. We thus learn
that it is impossible satisfy the strong form of Polchinski equation
without introducing multiple trace operators.

On the other hand, there is no problem to satisfy the weak form 
of Polchinski's equation, eq.\wjoe. It simply generates a 
first order differential equations for $\phi_n(t)$. Also, as we shall see
shortly, the term $\p^2 S_1$ in \wjoe\ is the same order as the
term $\p S_1\p S_1$, in the large N limit. This is quite different
from the speculation of \dvv, where it is conjectured that the holographic
RG equation is just the strong form of Polchinski equation, if so,
then $\p^2 S_1$ is suppressed by $1/N^2$.

Define the beta function
\eqn\betaf{\beta_n(\phi)={d\phi_n\over dt},}
then 
\eqn\fterm{\langle\p_t S_1\rangle=\beta_n N\langle\tr\Phi^n\rangle
=\beta_n{\p S\over \p\phi_n},}
where we suppressed summation over $n$. Use \oned,
\eqn\onedd{\langle {\p S_1\over\p\Phi_{ij}}{\p S_1\over\p\Phi_{ji}}
\rangle=Ng_n{\p S\over \p\phi_n}.}
Use \twod,
\eqn\twodd{\langle{\p^2S_1\over\p\Phi_{ij}\p\Phi_{ji}}\rangle
=N^{-1}G_{mn}e^{-S}{\p^2\over\p\phi_m\p\phi_n}e^S
=N^{-1}G_{mn}({\p S\over\p\phi_m}{\p S\over \p\phi_n}+{\p^2 S
\over \p\phi_m\p\phi_n}).}
Substituting \betaf, \onedd\ and \twodd\ into \wjoe, we find
\eqn\hrg{(\beta_n+\half\p_tK^{-1}g_n){\p S\over\p\phi_n}
+\half N^{-2}\p_t K^{-1}G_{mn}({\p S\over\p\phi_m}
{\p S\over\p\phi_n}+{\p^2 S\over \p\phi_m\p\phi_n})=0.}
This equation is almost the same as the holographic RG equation,
say as presented in \kv. The RG equation in \kv\ is derived for
the leading order in the large N limit. To compare with that,
we use the genus expansion of \gexp\ to derive in the leading order
\eqn\gez{(\beta_n+\half\p_tK^{-1}g_n){\p F_0\over\p\phi_n}
+\half\p_tK^{-1}G_{mn}{\p F_0\over\p\phi_m}{\p F_0\over\p\phi_n}=0.}

Two crucial points deserve mentioning explicitly. unlike one would
naively think, the term $\p S_1\p S_1$ in the weak form of Polchinski
equation is not identified with $\p S\p S$ in the holographic RG
equation. Rather, it generates only the form $\p S$, a correction to the beta
function in \hrg. On the other hand the term $\p^2 S_1$ in the
weak form of Polchinski equation generates both the terms $\p S\p S$ and
$\p^2 S$ in the holographic RG equation, with the term $\p^2 S$
subleading to $\p S\p S$ in the large N limit. Clearly, the former
is identified with the disconnected part of two point functions,
as also used in \kv, while the latter is identified with connected
part of two point functions. Clearly, both of these terms appear
only in a large N theory, since they come from the double traced operators
in the Polchinski equation. With a single scalar field, one has only
the $\p S$ term. The resulting RG equation can not be interpreted
as coming from the Hamilton-Jacobi equation of a gravity theory. 

In the AdS/CFT correspondence, the beta functions are determined by the
local part of the effective action $S$. Denote this local part by
$S_{loc}$. It contains kinetic term $\half G^{mn}\p_t\phi_m\p_t\phi_n$,
so the beta function is given by
\eqn\bdet{\beta_m=G_{mn}{\p S_{loc}\over\p \phi_n},}
where $S_{loc}$ is a functional of $\phi_m (t)$ which is determined
by the initial value problem.
However, in our toy model there is no such a kinetic term in the
extra dimension $t$, the reason is quite simple: the effective action
\effc\ is defined already as a functional of initial values $\phi_m(t)$.
We do not know how to define an ``off-shell" action which can be expressed
as an integral over the whole range of $t$. In fact, the beta functions
are not completely determined by demanding RG invariance of the effective
action only. 

The 0-dimensional one-matrix model can be solved completely in the large
N limit \mm. For instance, one can derive the Schwinger-Dyson equation
in this limit.   
Apparently given  a finite set of functions $\{\phi_n(t)
\}$, one of them is determined by the rest by requiring RG invariance.
Our above discussions serve only for the purpose of deriving the holographic
RG equation from the weak Polchinski equation. Eq.\hrg\ has a flavor
of the string equation and Virasoro constraints in one-matrix model.
Incidentally these equations can be derived by an action principle \ty.
It may be worthwhile to pursue along this direction.

The beta-functions are completely determined in a field theory. The
new ingredient here is the requirement of the cut-off independent
correlation functions. 
It is straightforward to generalize the above consideration to one
matrix model in D dimensional spacetime. We use Euclidean signature.
Now the regularized kinetic action is
\eqn\fki{S_0=-\half N\int d^Dxd^DyK(x-y,t)\tr\p_i\Phi(x)\p_i\Phi(y),}
where $K(x-y,t)$ is the cut-off inverse propagator. When $t\rightarrow
-\infty$, it tends to a delta function. Since we are dealing with
a field theory, in order to have a RG
invariant partition function, it is not enough to have a local interaction
action $S_1$. Assume the inverse of $K(x-y,t)$ exist (so that the
kinetic term is not degenerate), denote this inverse by $K^{-1}(x-y)$:
\eqn\invp{\int d^DzK^{-1}(x-z)K(z-y)=\delta^D(x-y).}
The weak form of Polchinski equation reads
\eqn\fwp{\langle \p_tS_1+N^{-1}\int d^Dxd^DyK_1(x-y)
\left({\p S_1\over\p \Phi(x)_{ij}}{\p S_1\over \p\Phi(y)_{ji}}
+{\p^2S_1\over\p\Phi(x)_{ij}\p\Phi(y)_{ji}}\right)\rangle =0,}
where
\eqn\kone{K_1(x-y)=-\half \Delta^{-1}\int d^Dx'd^Dy'  K^{-1}(x-x')\p_t
K(x'-y')K^{-1}(y'-y).}

In order to satisfy the above equation, $S_1$ must contain all the
nonlocal terms. If we start with a local action
\eqn\lac{S_1=N\sum\int d^Dx\phi_n(x)\tr\Phi^n(x),}
then
\eqn\nogen{{\p S_1\over\p \Phi(x)_{ij}}{\p S_1\over \p\Phi(y)_{ji}}
=N^2\sum\phi_m(x)\phi_n(y)\tr\Phi^m(x)\Phi^n(y),}
infinitely many nonlocal terms are generated. The above can be expanded
in derivatives of $\Phi$. On the other hand, 
\eqn\cont{{\p^2S_1\over\p\Phi(x)_{ij}\p\Phi(y)_{ji}}=
\sum N\delta^D(x-y)(m+n+2)\tr\Phi^m(x)\tr\Phi^n(y),}
yielding a contact term. This contributes to the holographic RG equation
a term 
\eqn\hrge{N^{-2}K_1(0)\sum (m+n+2)\int\phi_{m+n+2}(x)
\left({\p S\over\phi_m(x)}{\p S\over\p\phi_n(x)}+{\p^2S\over\p\phi_m(x)
\p\phi_n(x)}\right).}
This is good news, since in the holographic RG equation, this is indeed
a contact term. It originates from the fact that $\phi_m(x)$ is a field in the
AdS space, thus has a local quadratic term in the effective action. Denote
the Fourier transform of $K(x-y)$ by $K(p)$, then $K^{-1}(p)\rightarrow 0$
when $|p|\rightarrow \infty$, and the coefficient $K_1(0)$ in \hrge\ is given
by
\eqn\kzero{K_1(0)=\half\int d^Dp {1\over p^2}\p_t K^{-1}(p),}
which is certainly convergent.

To have a closed form of RG equation, we thus introduce all possible
single trace operators into the interaction part $S_1$. A generic operator
is
$$\tr\p_{i_1}\dots \p_{i_n}\Phi\dots \p_{j_1}\dots\p_{j_m}\Phi.$$
Denote such a generic operator by ${\cal O}_I(x)$, and the corresponding
coupling by $\phi^I(x)$.
Now go through the above steps, we will arrive at the following holographic
RG equation
\eqn\grg{\int d^Dx\left[(\beta^I+g^I){\p S\over \p\phi^I(x)}+G_{IJ}
\left({\p S\over\p\phi^I(x)}{\p S\over \p\phi^J(x)}+{\p^2 S\over
\p\phi^I(x)\p\phi^J(x)}\right)\right]=0,}
where all the components of the metric $G_{IJ}$ are proportional to an integral
of $p^{-2}\p_tK^{-1}(p)$ weighted by a polynomial of $p$.
So for each component
of $G_{IJ}$ to be well-defined, the cut-off propagator $K^{-1}(p)$ must fall
off more rapidly than any negative power of $p$ for large $|p|$.
Note again that the correction to the beta function, $g^I$, comes from
the part $\p S_1\p S_1$ in the weak Polchinski equation. To compare with
\kv, we can identify $\beta^I+g^I$ with the beta function defined on
the world-sheet, where the  cut-off is defined on the world-sheet. As
already pointed out in \kv, there is a UV/UV relation between string
world-sheet physics and spacetime physics. The two cut-offs are not
identical, thus the two definitions of the beta functions are not the
same.  Once again, our RG equation \grg\ is valid for any $N$. Use the genus
expansion \gexp, one recovers the RG equation of \refs{\dvv,\kv} in the large
N limit. The subleading term $\p^2 S$ in \grg\ is to be interpreted as coming
from quantum corrections in the AdS/CFT context.

We also want to emphasize the fact that our RG equation \grg\
involves a single integral. This is also true for the equation derived
in \kv. In order to remove this integral to obtain a local form, we need
to introduce position-dependent cut-off. This is also related to general
covariance in AdS/CFT. We leave a detailed discussion of this to
another work.

More equations can be derived by demanding the renormalized correlation
functions to be independent of the cut-off. These equations are just
Callan-Symanzik equations. They can be derived from the
local form of the RG equation \dvv, but not from \grg. We can derive
them in the matrix field theory by generalizing the steps leading to \grg. 
These equations
put together will give a closed system of equations for $S$ and $\beta^I$.
We are not sure whether these beta functions can be written in a form
\bdet. Note that the metric $G_{IJ}$ in \bdet\ on the moduli space 
is the same as in \grg, and is already determined in deriving \grg.
It would be highly nontrivial if all these beta functions are given
by a single functional $S_{loc}$. Maybe this is the most crucial
criteria for a holographic theory, and is generically violated by an
arbitrary matrix field theory such as the single scalar field theory.

It remains to generalize our construction to ${\cal N}=4$ super Yang-Mills
theory. Although we do not see essential difficulty in doing this,
we need to resolve the problem of introducing a gauge invariant cut-off.
A naive cut-off will not work, since this is not compatible with
local gauge transformation which mixes all energy scales. Another way
to see this is through the naive cut-off Yang-Mills action 
$$\int d^Dx \tr F_{\mu\nu}(x)F_{\mu\nu}(y) K(x-y,t),$$
it is certainly not gauge invariant. It has been suggested to use Wilson
loop as gauge invariant variable to overcome this difficulty \morris,
and this line of approach was followed up in \hirano. However, the existence
of AdS/CFT correspondence indicates that a gauge invariant cut-off
exists for local gauge invariant variables. We suspect that stochastic
quantization \pw\ may be one way to gauge invariantly regulate 
Yang-Mills theory. And the stochastic time has been interpreted as the 
RG scale recently in \vipul.

The discussion presented here may be viewed as a zero-slope limit of
approach of \kv. However, the relation between string world-sheet and
large N diagrams need to be further clarified. Also, our approach does not
has the drawback of assuming perturbation theory as in \kv. The open/closed
string duality need to be understood. One particularly nice example was
discussed in \lw.

Acknowledgments. This work was supported by a grant of NSC and by a 
``Hundred People Project'' grant of Academia Sinica. I thank M. Yu
for several useful conversations, and T. Yoneya for comments on the
manuscript.

\vfill
\eject

\listrefs
\end